\documentclass[11pt]{article}
\pdfoutput=1
 \usepackage{mciteplus}
 \usepackage{tikz}
 \usepackage{color}
 \usepackage{xcolor}
 \usepackage{comment}
 \definecolor{darkblue}{rgb}{0.1,0.1,.7}
 \usepackage[colorlinks, linkcolor=darkblue, citecolor=darkblue, urlcolor=darkblue, linktocpage,hyperfootnotes=false]{hyperref} 
\usepackage{epsfig}
\usepackage{graphicx}
\usepackage{cite}
\usepackage{amsfonts}
\usepackage{amssymb}
\usepackage{bm}
\usepackage{latexsym}
\usepackage{mathtools}
\usepackage{amsmath}
\numberwithin{equation}{section}
\def\bq{\begin{quote}}
\def\eq{\end{quote}}
 at 10truept

\newcommand{\calc}{{\cal C}}
\newcommand{\calo}{{\cal O}}
\newcommand{\calh}{{\cal H}}

\newcommand{\call}{{\cal L}}
\newcommand{\calp}{{\cal P}}

\newcommand{\beq}{\begin{equation}}
\newcommand{\eeq}{\end{equation}}
\newcommand{\beqa}{\begin{eqnarray}}
\newcommand{\eeqa}{\end{eqnarray}}
\newcommand{\bea}{\begin{eqnarray}}
\newcommand{\eea}{\end{eqnarray}}

\newcommand{\hf}{\frac{1}{2}}

\def\roughly#1{\raise.3ex\hbox{$#1$\kern-.75em\lower1ex\hbox{$\sim$}}}

\begin{document}

\thispagestyle{empty}
\begin{titlepage}
  \bigskip

  \bigskip\bigskip

  \bigskip

\begin{center}
{\Large \bf {Perturbative
quantum evolution of the gravitational state and dressing in general backgrounds
}}
    \bigskip
\bigskip
\end{center}

  \begin{center}

 \rm {Steven B. Giddings\footnote{\texttt{giddings@ucsb.edu}} and Julie Perkins\footnote{\texttt{jnperkins@ucsb.edu}}}
  \bigskip \rm
\bigskip

{Department of Physics, University of California, Santa Barbara, CA 93106, USA}  \\
\rm

  \bigskip \rm
\bigskip
 
\rm

\bigskip
\bigskip
\begin{center}
{\sl Dedicated to the memory of Jim Hartle}
\end{center}

  \end{center}

\vspace{3cm}
  \begin{abstract}

This paper sets up a perturbative treatment of the evolving quantum state of a gravitational system, in a Schr\"odinger-like picture, working about a general background. This connects gauge symmetry, the constraints, gravitational dressing, and evolution.  Starting with a general time slicing, we give a simple derivation of the relation between the constraints, the hamiltonian, and its well-known boundary term.  Among different approaches to quantization with constraints, we focus on a ``gauge-invariant canonical quantization," which is developed perturbatively in the gravitational coupling.  The leading-order solution of the constraints (including the  Wheeler-DeWitt equation) for perturbations about the background is given in terms of an explicit construction of gravitational dressings built using certain generalized Green functions; different such dressings corresponding to adding propagating gravitational waves to a particular solution of the constraints.  Dressed operators commute with the constraints, expressing their gauge invariance, and have an algebraic structure differing significantly from the undressed operators of the underlying field theory.  These operators can act on the vacuum to create dressed states, and evolution of general such states is then generated by the boundary hamiltonian, and alternately may be characterized using other relational observables.  This provides a concrete approach to studying perturbative time evolution, including the leading gravitational backreaction, of quantum states of black holes with flat or anti de Sitter asymptotics, for example on horizon-crossing slices.  This description of evolution in turn provides a starting point for investigating possibly important corrections to quantum evolution, that go beyond quantized general relativity.

 \medskip
  \noindent
  \end{abstract}
\bigskip \bigskip \bigskip 

  \end{titlepage}

\section{Introduction and motivation}

If there is a quantum-mechanical theory of gravity, the big challenges in its formulation include understanding the fundamental description of its quantum states and observables, as well as the nature of the unitary evolution on its Hilbert space.  Approaches to this problem based on quantizing general relativity (GR) or related classical theories have run into vexing problems, initially nonrenormalizability\footnote{For a review with further references, see \cite{Burg}.} but likely more profoundly that of nonunitarity in the high-energy regime involving black holes.  An alternative approach is to begin with the hypothesis that one is working with a quantum-mechanical theory, and investigate what mathematical structure of such a theory is necessary to describe gravity and consistently match the known and tested physics of local quantum fields propagating on a weakly-curved background, in the appropriate limits.  This might be referred to as a ``quantum-first" approach \cite{UQM,SGAlg,QFG,QGQF}.\footnote{For related discussion, see \cite{CCM,CaCa,CaSi}.  Also note that if one can find a complete ``holographic map," the AdS/CFT correspondence could be an approach to providing such structure.}  One would like to understand the nature of the Hilbert space for gravity, and of its algebras of observables, symmetries, and unitary evolution law.

Such an approach does not argue for completely abandoning a perturbative quantization of GR.  The match to the known and tested physics of local quantum field theory (LQFT) on a weakly curved background, the confirmed existence of gravitational waves, and the apparent approximate validity of strong-field classical solutions suggests that such a perturbative treatment gives at least {\it approximately} correct physics, in the weak-gravity regime, though one that is missing important effects in other regimes.  One can view this as a ``correspondence principle" for quantum gravity.  What is interesting is that already in this limit, one encounters non-trivial new properties of quantum gravity that signal its departure from LQFT.  This also raises the hope that, by better understanding this structure in the perturbative limit, one may infer key properties of the more basic structure of a fundamental theory of quantum gravity.

In particular, a significant part of the difficulty of gravity seems to stem from the form of its gauge symmetries.  And, significant aspects of this non-trivial gauge structure appear to already be present at leading perturbative orders.  This suggests that a useful starting point is simply to better understand this structure at these leading orders.

One aspect of this structure is the lack of local gauge invariant observables\cite{Torr}.  In short, any local observable clearly carries nontrivial Poincar\'e charge (in the example of flat asymptotics), since it doesn't commute with translation generators, and this must source an associated gravitational field that extends to infinity\cite{DoGi2}.  

There are different approaches to constructing {\it non}local observables that respect gauge invariance, typically in a ``relational" fashion.  One approach is to specify the position of a quantum operator relative to other quantum fields that vary in spacetime; we refer to the resulting operator as ``field-relational" (or, observer-relational), and an example is provided by calculation of primordial perturbations in cosmology by referring to the time of reheating set by the inflaton in inflationary models.  Another alternative is to construct relational observables by using position information from the gravitational sector; perturbatively, one can begin with a local observable, and ``gravitationally dress" it to construct gauge-invariant operators that no longer commute at spacelike separation\cite{SGAlg}\cite{DoGi1}\cite{DoGi2}\cite{DoGi3}\cite{QGQF}\cite{DoGi4,GiKi}.\footnote{To clarify a difference in terminology, the recent work
\cite{CLPW} considers observables referred to an observer but calls them dressed, despite not having a gravitational component; here those would be referred to as field-relational.  Earlier work related to dressed observables includes \cite{Heem} and \cite{KaLigrav}; the first derived nontrivial commutators as arising from constraints, but didn't give the dressed operators, and the second focussed on deriving {\it commuting} operators. For related constructions in the cosmological setting, see \cite{Brunetti:2016hgw}.}
In quantization of GR, the diffeomorphism symmetry is generated by the gravitational constraints, so in either approach, a test for gauge invariance of such operators is that they commute with the constraints.

Another key question for a quantum-mechanical theory is the structure of its states and their evolution.  The observables both characterize states, and furnish a means of constructing gauge-invariant states: one can act with a quantum observable on a ``vacuum" state to create a nontrivial state.  The important question of evolution of these states, if it is unitary, can then be addressed by providing a hamiltonian.\footnote{Even in LQFT, it is argued that the hamiltonian provides a more  fundamental description of evolution than an action; see {\it e.g.} \cite{Ramo}, sec.~2.2.}  In LQFT coupled to quantized GR, as we will review and further clarify, the hamiltonian and evolution are of course closely related to the constraints.  In a closed universe, the hamiltonian is given by the constraints, and so formally vanishes on their solution; in a universe with asymptotic spatial infinity, there is an additional term in the hamiltonian, that is important for evolution.   In particular, solving the quantum version of the hamiltonian constraint is commonly referred to as solving the Wheeler-de Witt (WdW) equation, and is accomplished by gravitationally dressing undressed operators or states.\footnote{Ref.~\cite{Rajuetal} also discusses perturbative solution of the WdW equation, but seems not to have realized that this is achieved by constructing gravitational dressing, nor recognized the relevance of preceding works on this subject.}
 An additional subtlety (see below) is that physical states may only be annihilated by ``half" of these constraints.  The form of the evolution in the perturbative regime is expected to furnish clues about its nonperturbative completion.

Construction of the gravitational dressing, which can be explicitly treated at leading order in the gravitational coupling, is also relevant to the question of holography of gravity.  A leading proposed explanation of holography in anti de Sitter space (AdS) is that it follows from the hamiltonian being a boundary term\cite{Maroholo,Maroholonost} when the constraints are satisfied; for additional discussion, see \cite{Jacoholo,SGholo}. A closely similar argument is that momentum generators are also boundary terms, and so one can act both to translate a state to infinity, and to measure it, purely with operators at infinity\cite{DoGi3}.  There are related arguments that even perturbative observables at infinity can determine bulk states\cite{Rajuetal}, but their sensitivity is exponentially suppressed\cite{SGsub}. These statements do rely on solution of the constraints, so implicitly on solution of the bulk dynamics\cite{SGholo}.\footnote{Entanglement wedge reconstruction appears to likewise assume solution of the constraints\cite{JLMS}\cite{SGholo}.}   

However these arguments are ultimately understood, it is clear that the dressing modifies the locality properties of the algebra of operators\cite{SGAlg,DoGi1}; it also apparently connects\cite{SGinpro} to recent discussion of modification of the structure of von Neumann algebras from type III to II\cite{Wittcross}.

In short, at the perturbative level it appears that we can begin to learn important aspects of the structure of observables, states, and their evolution.  This description is of course expected to miss crucial effects, particularly when treating strong gravitational configurations such as black holes.\footnote{For an approach to parameterizing such effects as departures from LQFT evolution, see \cite{NVU,BHQU} and references therein.}  But, a clearer understanding of the perturbative structure is also expected to help provide a basis and background for understanding the role of modifications in the strong gravity context, if the complete theory respects the correspondence principle and is consistent with its weak gravity limit.  

In the interest of such a deeper understanding of the interplay of evolution, gauge symmetry, the constraints, and gravitational dressing, in general contexts, this paper will investigate the perturbative structure of the hamiltonian and constraints, working perturbatively about a general background.  The next section begins with a simplified derivation of the hamiltonian, exhibiting it either in a more conventional local form familiar from LQFT, or as a term proportional to the constraints plus a boundary term.  Section three then outlines different approaches to perturbative quantization, and sets up a perturbative treatment of  the constraints in what we refer to as ``gauge-invariant canonical quantization."  Section four gives a leading perturbative construction of operators commuting with the constraints, working about a general background, in terms of a construction of the gravitational dressing that generalizes \cite{DoGi1}\cite{QGQF}\cite{DoGi4,GiKi,SGsplit}.  Section five discusses construction of corresponding states, briefly discusses the form and characterization of their evolution, and illustrates application to the important cases of black holes and/or AdS spacetimes.  Section six finishes with conclusions and further directions.
Appendices illustrate basic features of the analogous treatment of electromagnetism, and show how the constraints generate gauge transformations.

\section{Action, hamiltonian, and boundary terms}

\subsection{Action and boundary terms}

This section will review formulation of the action in Arnowitt-Deser-Misner (ADM)\cite{ADM} variables, and describe a  simple approach to deriving the appropriate boundary terms.  This  paper focusses on quantization of Einstein gravity plus matter, perturbing around a general background.  The usual starting point is the action\footnote{We find it most convenient to work with expressions for lagrangians and hamiltonians that are  densities.}
\beq\label{action}
S=\int d^Dx \left( \frac{1}{16\pi G} \sqrt{|g|}R + \call_m\right) + S_\partial
\eeq
in $D$ spacetime dimensions, where $G$ is Newton's constant, $\call_m$ is a matter lagrangian, and $S_\partial$ is a boundary term.  If a specific matter action is needed, the scalar theory with
\beq\label{scalarlagrange}
\call_m= -\sqrt{|g|}\left[\frac{1}{2} (\nabla \phi)^2+V(\phi)\right]
\eeq
furnishes a useful example.   

\begin{figure}[!hbtp] \begin{center}
\includegraphics[width=15cm]{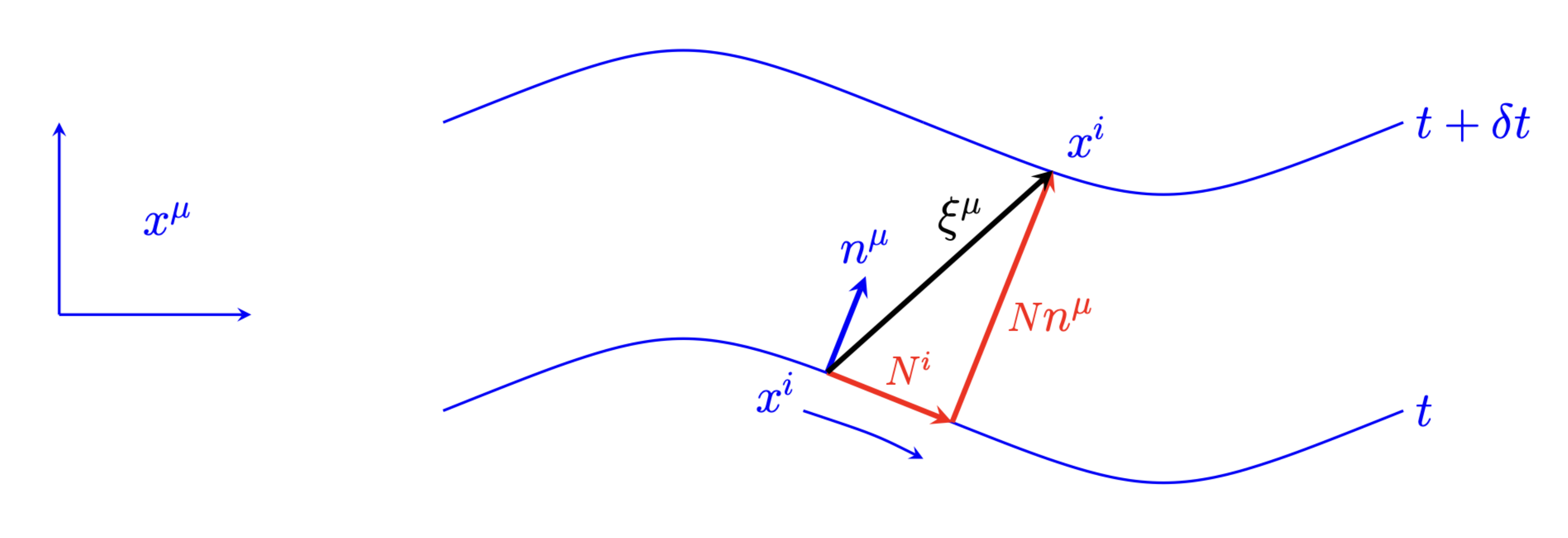}
\end{center}
\caption{Shown are two members of a family of slices labelled by $t$.  Points at the same spatial coordinate $x^i$ are connected by the vector $\xi^\mu$, which can be decomposed in terms of normal and tangential components to give the lapse and shift; vectors in the figure are scaled by an implicit $\delta t$.}
\label{Figslices}
\end{figure}

Since our focus will be on the evolving quantum state describing perturbations about a background, we would like to find a corresponding hamiltonian.  We begin by 
introducing a foliation of the spacetime by slices labelled by time $t$, and with spatial coordinate $x^i$, with relation to general coordinates given by
\beq
x^\mu= {\cal X}^\mu(t,x^i)\ .
\eeq
The displacement vector between points of equal $x^i$ on nearby slices is given by $\xi^\mu = (\partial {\cal X}^\mu/\partial t)_{x^i}$, and can be decomposed into pieces normal and tangential to a slice, 
\beq\label{xiton}
\xi^\mu=N n^\mu + N^\mu\ ,
\eeq
where $N$ is the lapse, $N^\mu$ is the shift, and $n^\mu$ is the unit normal; in $(t,x^i)$ coordinates $N^\mu=(0,N^i)$, and
these quantities are illustrated in Fig.~\ref{Figslices}.
In the coordinates $(t,x^i)$ of the foliation the metric takes the ADM form
\beq
ds^2=-N^2 dt^2 +q_{ij}(dx^i+N^i dt) (dx^j+N^j dt)\ 
\eeq
and the  unit normal to the slices has components
\beq
n^\mu=\frac{1}{N} (1,-N^i)\ .
\eeq

The gravitational lagrangian can then be derived in these variables, after introducing the extrinsic curvature of the slices,\footnote{Note that there are differing sign conventions in the literature; {\it e.g.} \cite{Waldbook,kiefer} differ by a sign.}
\beq\label{excurv}
K_{ij}= \frac{1}{2N}\left(-\dot q_{ij} + D_iN_j + D_jN_i\right)\ ,
\eeq
with dot denoting $\partial/\partial t$, $D_i$ the covariant derivative constructed from $q$, and latin indices raised/lowered with the spatial metric $q$.  This lagrangian is given by\cite{KuchII,KRV}
\beq
 \sqrt{|g|}R = N\sqrt q \left[\left(K_{ij}K^{ij} - K^2\right) + R_q\right] - 2\partial_i\left(\sqrt q q^{ij} \partial_j N\right) + 2\partial_i\left(\sqrt q KN^i\right) -2\partial_t(\sqrt q K)\ ,
 \eeq
with $K=q^{ij}K_{ij}$ and $R_q$ the scalar curvature of $q$.  
The total derivative terms become boundary terms in the action, which can be cancelled by $S_\partial$,
\beq
S_\partial = \oint \frac{dt dA^i}{8\pi G} (\partial_i N - K N_i) + \int d^Dx \frac{\partial_t(\sqrt q K)}{8\pi G} + S_\partial'\ .
\eeq
An additional term $S_\partial'$ is required, as is argued by requiring a well-defined variational principle in \cite{ReTebdy} or finiteness of the action in \cite{HaHo}.  
This can be described by introducing a background metric $g_0$, {\it e.g.} the Minkowski or anti de Sitter metrics, depending on boundary conditions, and writing the full metric as
\beq\label{metexp}
g_{\mu\nu} = g_{0\mu\nu} + \Delta g_{\mu\nu}\quad ,\quad q_{ij}= q_{0ij} +\Delta q_{ij}\ ;
\eeq
 it then takes the form
\beq\label{Spp}
S_\partial'= - \oint \frac{dt dA^i}{16\pi G} N \left( D_0^j \Delta q_{ij} - \partial_i \Delta q\right)\ ,
\eeq
with $dA^i$ the area element and $\Delta q = q_0^{ij} \Delta q_{ij}$.  This can be checked (using  equation \eqref{hlinvar} below) to eliminate the problematic boundary terms in the variation of the action.

The resulting gravitational action can be written in terms of a local lagrangian,
\beq
S_g = \int d^D x \call_g = \int d^D x \frac{N\sqrt q}{16\pi G} \left[(K_{ij} K^{ij} - K^2) + R_q\right] + S_\partial'\ ,
\eeq
with $S'_\partial$ rewritten as a volume integral.
The structure of $\call_g$ can for example be investigated by using the 
relation\cite{Edd,Sch,PMB}
\beq\label{indrel}
\sqrt q R_q =\sqrt q q^{lm}\left(\gamma^i_{jl} \gamma^j_{im} - \gamma^i_{lm} \gamma^j_{ij} \right) + \partial_i\left[\sqrt q \left(q^{jk}\gamma^i_{jk} - q^{ij} \gamma^k_{jk}\right)\right]\ ,
\eeq
with $\gamma^i_{jk}$ denoting the Christoffel symbols computed from the metric $q$.  If the metric is expanded about a background solution as in \eqref{metexp}, the linear terms vanish by the equations of motion of the background or cancellation with the boundary term, and quadratic and higher-order terms in the expansion of \eqref{indrel} give a lagrangian with quadratic contributions of the form $(\partial \Delta q)^2$, plus interaction terms.

 \subsection{Hamiltonian and constraints}
 
 Momenta conjugate to the spatial metric $q$ are defined as
 \beq\label{conjmom}
 P^{ij} = \frac{\delta S_g}{\delta \dot q_{ij}} = -\frac{\sqrt q}{16\pi G}\left(K^{ij}- q^{ij} K\right)\ ;
 \eeq
 we find it easiest to work with the form of these which are tensor {\it densities}.  The momenta conjugate to $N, N_i$ of course vanish, corresponding to the fact that the lapse and shift are Lagrange multipliers enforcing constraints.  The gravitational  action can then be rewritten in the canonical form
\beq
S_g = \int d^D x \left( P^{ij}  \dot q_{ij} -  \calh_g\right)\ .
\eeq
The hamiltonian  is found by a straightforward calculation to be
\bea\label{hamdens}
\int dt H_g &=& \int d^{D} x 
 \calh_g\ = \int d^{D} x (P^{ij}  \dot q_{ij} -\call_g)\cr &=& \int d^{D} x\left[ \frac{16\pi G  N}{\sqrt q} \left( P^{ij} P_{ij} -\frac{P^2}{D-2}\right) - \frac{N\sqrt q}{16\pi G}R_q +  2 P^{ij} D_i N_j \right] - S_\partial'\ .
 \eea

The hamiltonian can be rewritten in  different ways.  First, as expected from the description of $S_g$ given above, the expression $\calh_g$ in  \eqref{hamdens} is quadratic in momenta and first derivatives of the metric perturbation.  Alternately,  \eqref{hamdens} can be rewritten in terms of the Einstein tensor as
\beq\label{HamEin}
 \int dt H_g=\int d^{D}x\left[  -\frac{\sqrt q G_{nt}}{8\pi G} + 2D_i(P^{ij} N_j)\right] -S_\partial' ,
\eeq
where $G_{nt} = n^\mu \xi^\nu G_{\mu\nu}$.  The matter hamiltonian likewise is given in terms of the stress tensor
\beq\label{phistress}
T_{\mu\nu}=\nabla_\mu\phi \nabla_\nu\phi - \hf g_{\mu\nu} (\nabla \phi)^2
\eeq
as
\beq
\label{matham}
 \calh_m = \sqrt q T_{nt}=\sqrt q \frac{N}{2} \left(\frac{\Pi^2}{q} + q^{ij} \partial_i\phi\partial_j\phi\right) + \Pi N^i \partial_i\phi\ ,
\eeq
where the densitized canonical momentum is $\Pi=\sqrt q \partial_n\phi$.
Then, the full hamiltonian becomes
\beq\label{htot}
H = \int d^{D-1} x\, \calc_\xi + H_\partial\ ,
\eeq
with 
\beq
\calc_\xi\coloneqq  \xi^\mu \calc_\mu \coloneqq \xi^\mu\sqrt q\left( -\frac{G_{\mu\nu}}{8\pi G} + T_{\mu\nu} \right) n^\nu
\eeq
giving the usual gravitational constraints.  The boundary contribution is, from \eqref{Spp} and \eqref{HamEin},
\beq\label{boundham}
H_\partial= \oint dA^i \left[\frac{N}{16\pi G} \left(D_0^j \Delta q_{ij} -\partial_i \Delta q\right) + 2 \frac{P^{ij}}{ \sqrt q} N_j \right]\ ,
\eeq
which is the expected boundary expression
for gravity\cite{ReTebdy}, $-N(\infty)P_0^{ADM}-N^i(\infty) P_i^{ADM}$. As is well known, if the constraints are satisfied,
\beq\label{constraints}
\calc_\mu=0\ ,
\eeq
then the hamiltonian becomes simply this boundary expression \eqref{boundham}.

The interplay of the expressions \eqref{hamdens} and \eqref{htot} is worth noting, and can be summarized in
\beq\label{hreln}
H = \int d^{D-1}x  \left(N \calc_n + N^i \calc_i\right) + H_\partial = \int d^{D-1}x \, (\calh_g +\calh_m)
\eeq
where we have used  \eqref{xiton} relating $\xi$ to $n$, and define $\calc_n=n^\mu \calc_\mu$.   
On the one hand, using the expansion of $\call_g$ described in connection with \eqref{indrel},
 the rightmost expression in \eqref{hreln}
is of the general expected form for a field theory, with quadratic terms in momenta and derivatives of fields, as well as interaction terms.  We can think of this as generating time evolution in the usual way.  On the other hand, one can work with a solution of the constraints \eqref{constraints}, in which case the hamiltonian reduces to the surface term $H_\partial$.  The later observation has been argued to be connected to the ``holographic" property of gravity \cite{Maroholo,Maroholonost,Jacoholo}, but does rely\cite{SGholo} on first solving the constraints, which behave as equations of motion.  

An important question is thus the role of the constraints and different forms for the hamiltonian in the quantum theory, as well as their possible corrections from a more complete quantum theory.

\section{Quantization and perturbative expansion}

\subsection{Quantization, constraints, and gauge invariance}\label{quantization}

Our goal is to find a consistent quantum theory reproducing the preceding classical structure in the appropriate limits.\footnote{For an analogous discussion for QED, see Appendix \ref{appa}.}  The canonical approach tells us to introduce canonical commutators,
\beq\label{cancom}
[P^{ij}(x,t),q_{kl}(x',t)] = -i \delta^i_{(k}\delta^j_{l)} \delta^{D-1}(x-x')\ ,
\eeq
with normalization 
\beq
\delta^i_{(k}\delta^j_{l)} = \hf\left( \delta^i_{k}\delta^j_{l} + \delta^i_l\delta^j_{k}\right)\ .
\eeq
Since $N$ and $N^i$ have vanishing conjugate momenta, they are taken to be c-numbers,\footnote{Here we work on a ``reduced" phase space; for further discussion see comments in Appendix  \ref{appb}, and {\it e.g.} \cite{RTH} or \cite{PSS}.} and \eqref{hreln} shows their role as Lagrange multipliers for the constraints.  This means that these variables are not determined by the equations of motion, and their arbitrariness is part of the gauge symmetry.  Gauge transformations acting on the canonical variables $(q_{ij},P^{ij})$ (and $(\phi,\Pi)$) are generated by $\calc_n$ and $\calc_i$, as described in Appendix \ref{appb}.

The Heisenberg equations of motion take the form 
\beq\label{qeqn}
\partial_t q_{ij} = i [H,q_{ij}]\ ,
\eeq
which reproduces the expression \eqref{excurv} for the extrinsic curvature, using \eqref{hamdens}, and 
\beq\label{Peqn}
\partial_t P^{ij} = i [H,P^{ij}]
\eeq
which gives the $ij$ Einstein equations.  These time derivatives are not gauge invariant, unlike the case of QED (see Appendix \ref{appa}), since the gauge transformations act non-trivially on $q_{ij}$ and $P^{ij}$.  

The next question is how to describe physical states.  It is tempting to require $\calc_\mu|\psi\rangle=0$ for physical states, but this would then imply that for general operators $O$
\beq
\langle \psi| [\calc_\mu,O]|\psi\rangle =0\,
\eeq
conflicting with the preceding equations of motion.  Thus, in order to correctly describe nontrivial evolution, the constraints should not be taken to vanish identically on the physical Hilbert space.

Multiple related ways to proceed have been studied in the literature, in each of which the question of locality becomes nontrivial.  A brief summary is:

\begin{enumerate}

\item Dirac quantization.  Here one introduces gauge-fixing conditions and solves these and the constraints, and also introduces a new Dirac bracket (or alternately redefines operators) such that $[\calc_\mu,O]_D=0$.  This appears to simplify commutators, but it is also true that solving the gauge conditions and constraints is nonlocal.  This nonlocality is then ``hidden" in the structure of the Dirac brackets; a simple example of this for QED is described in sec. IV.A.3 of \cite{DoGi1}.  

\item Covariant gauge ``fixing" (breaking).  In this approach a gauge-violating term is added to the action, and then canonical commutators postulated for all components of the metric.  This was used to study gauge-invariant operators in \cite{DoGi1,GiKi}; the gauge breaking term decouples for these.  These operators are in general nonlocal, due to gravitational dressing, which is found by requiring vanishing commutators with the constraints\cite{DoGi2}.

\item BRST/BFV quantization\cite{FrVi,BaVi}.  Here extra fields, including ghosts, are added; extra conditions are necessary as well.

\item Refined algebraic quantization, in which group averaging of states on an auxiliary kinematic space induces an inner product on the space of states satisfying the constraints\cite{ALMMT,MaroRAQ,GiuMa,Marogroupav}.

\item ``Gauge invariant canonical quantization." 

\end{enumerate}

The latter approach appears to be distinct from approaches previously described in the literature; it is briefly described for QED in the Appendix \ref{appa}, and will be utilized here.  While it is closely similar to covariant gauge breaking used in \cite{DoGi1,GiKi}, the constraints are separated into what may be called the positive and negative frequency parts, and the positive frequency constraint is taken to annihilate the vacuum state. Then, the constraints are found to commute with the gauge-invariant operators as usual, and also define the time evolution of the quantum state via the Hamiltonian.  Specifically, one assumes a suitable decomposition of the constraints
\beq\label{Consdecomp}
\calc_\mu(x)= \calc_\mu^+(x) + \calc_\mu^-(x)
\eeq
into positive and negative frequency parts, and then imposes the physical state condition
\beq\label{physstate}
\calc_\mu^+(x)|\psi\rangle=0
\eeq
for a $|\psi\rangle$ taken to be the vacuum state $|0\rangle$, on a given time slice.  Since the constraints generate gauge transformations (see Appendix \ref{appb}), gauge-invariant operators are those satisfying
\beq\label{conscomm}
[\calc_\mu(x),O]=0\ .
\eeq
Such an operator creates a non-trivial state, $|\psi\rangle = O |0\rangle$, which also satisfies \eqref{physstate} by virtue of $[\calc_\mu^+,O]=0$.  

A decomposition \eqref{Consdecomp} is possible for example when perturbing about stationary backgrounds, using the corresponding Killing vector to define frequency.  For perturbations about time-dependent backgrounds, there are additional complications\cite{ToVa,CMOV,CFMM,AgAs,MuOe} with such a picture which we will not address here.  However, we note that the condition for gauge-invariant operators, \eqref{conscomm}, is independent of this decomposition, and so that should not play a direct role in their construction.

Suppose we consider evolution of states, in a Schr\"odinger picture, via the hamiltonian  \eqref{hreln}.
Since the full constraints don't annihilate physical states, the  time dependence of states will depend on gauge (here, choice of arbitrary ${\bf N}= (N, N^i)$); the gauge-dependent part of the change in the state for evolution for time $\delta t$ via the hamiltonian \eqref{hreln} is
\beq
\delta_{\bf N} |\psi\rangle = i \delta t \int d^{D-1}x  \left(N \calc_n^- + N^i \calc_i^-\right)|\psi\rangle\ .
\eeq
However, this will be orthogonal to another physical state $|\psi'\rangle$.   Likewise, if we consider evolution of a matrix element of a gauge invariant operator $O$,
\beq
\partial_t \langle \psi'|O |\psi\rangle = i\langle \psi'|[H,O]|\psi\rangle\ ,
\eeq
we find that this is also independent of the gauge-variant ($\bf N$-dependent) part of the hamiltonian \eqref{hreln}.  In short, while 
there is a gauge ambiguity in the states, that is not present in matrix elements of gauge-invariant operators.  Of course, we find from the Heisenberg equations \eqref{qeqn} and \eqref{Peqn} that evolution of matrix elements of $q_{ij}$ and $P^{ij}$ is gauge dependent.

\subsection{Perturbative expansion}

The remainder of this paper will primarily focus on a perturbative construction of states and operators like we have just described.  Some analogous work has been done in the cosmological setting, including expanding the constraints to second order and related methods to find gauge invariant observables \cite{langlois1994hamiltonian, noh2004second}. However, they do not consider general backgrounds nor the general construction of gravitational dressing. The ADM decomposition and  perturbation of the constraints has also been investigated in other special backgrounds, particularly  asymptotically flat \cite{kuchavr1970ground}, AdS \cite{chowdhury2022holography}, and spacetimes with black holes \cite{hollands2013stability}.  Going beyond this work, we will consider a perturbative expansion of the ADM decomposition on an arbitrary classical background, and use the new expansions of the constraints to define the dressed operators to leading order by requiring that the operators commute with the constraints, as will be described in the subsequent sections. We begin with a classical metric $g_{\mu\nu}\leftrightarrow(N,N_i,q_{ij})$ satisfying Einstein's equations, including the constraints \eqref{constraints}, possibly also with the stress tensor of a classical matter background $\phi_0$.  The corresponding quantum variables are denoted $\tilde g_{\mu\nu}\leftrightarrow(\tilde N,\tilde N_i,\tilde q_{ij})$ and $\tilde \phi$.  Introducing the parameter $\kappa^2=32 \pi G$, these may be expanded as
\beq
\tilde g_{\mu\nu} = g_{\mu\nu} + \kappa h_{\mu\nu}\ ,\ {\tilde \phi}=\phi_0 +\phi\ ,
\eeq
and likewise for $(\tilde N,\tilde N_i,\tilde q_{ij})$, in particular with
\beq
\tilde q_{ij}= q_{ij} + \kappa h_{ij}\ .
\eeq
We will also expand $\tilde P^{ij}$ as
\beq
\tilde P^{ij} = P^{ij} + \frac{p^{ij}}{\kappa}\ ,
\eeq
in which case the canonical commutators \eqref{cancom} also take the simple form
\beq\label{pertcom}
[p^{ij}(x,t),h_{kl}(x',t)] = -i \delta^i_{(k}\delta^j_{l)} \delta^{D-1}(x-x')\ .
\eeq
Notice, from \eqref{conjmom}, that for a non-trivial classical background, $P^{ij}\propto 1/\kappa^2$.  We also expand $\tilde \Pi = \Pi_0 +\Pi$.

The explicit form of the constraints (temporarily written without tildes) is
\beq\label{hamcons}
0=\calc_n=\sqrt q\left(T_{nn} - \frac{4}{\kappa^2} G_{nn}\right)
\eeq
and
\beq\label{momcons}
0=\calc_i =\sqrt q\left(T_{ni} - \frac{4}{\kappa^2} G_{ni}\right)\ .
\eeq
Here the pertinent components of the Einstein tensor are
\beq
- \frac{4}{\kappa^2} G_{nn} = - \frac{2}{\kappa^2}R_q + \frac{\kappa^2}{2 q} \left( P^{ij} P_{ij} -\frac{P^2}{D-2}\right)
\eeq
and
\beq
 - \frac{4}{\kappa^2} \sqrt q G_{ni} = -2 D_j P_i^{j}\ ,
 \eeq
 and those of the stress tensor are, from \eqref{matham}, 
\beq
\sqrt q T_{nn} = \frac{\sqrt q}{2} \left(\frac{\Pi^2}{q} + q^{ij} \partial_i\phi \partial_j\phi\right)\quad,\quad \sqrt q T_{ni} =\Pi \partial_i\phi\ .
\eeq
The expansions of the constraints \eqref{hamcons} and \eqref{momcons} must then be found for the quantum perturbations $h_{ij}$, $p^{ij}$ and $\phi,\Pi$.

The hamiltonian constraint has the expansion
\beq\label{Hamexp}
{\tilde \calc}_n =\calc_n + \hf q^{ij} \kappa h_{ij} \calc_n - \frac{4}{\kappa^2} \sqrt q \delta_{\kappa h}G_{nn} + \sqrt q\left(\delta_\phi T_{nn} +\delta_{\kappa h} T_{nn}\right) + \sqrt q T_{nn}^Q + \sqrt q \breve t_{nn}\ .
\eeq
Here $\delta_{\kappa h}$ and $\delta_\phi$ denote first order variations, and 
 we define a quantum stress tensor $T^Q$ which collects terms that are quadratic and higher order in the variables $\phi$, $\kappa h$, coming from $T_{nn}$, as well as a gravitational stress tensor $\breve t$ that contains the quadratic and higher order terms in $\kappa h$ and $p$ arising from the Einstein tensor term.  Since the background satisfies Einstein's equations, the first two terms vanish.  To find the third term and $\breve t$ we need the expansion of the Ricci scalar
\beq
R_{\tilde q} = R_q + \kappa \delta_h R_q + \delta^{\geq 2}_{\kappa h} R_q\ ,
\eeq
where the last term summarizes all higher-order terms in $\kappa h$.  Explicitly, the expansion of the Ricci scalar is well known
\beq\label{hlinvar}
\delta_h R_q = D^i D^j h_{ij} -R_q^{ij}h_{ij} -D_iD^i(q^{kl} h_{kl}) :=L^{ij} h_{ij}\ ,
\eeq
defining the second-order differential operator $L^{ij}$.  We also need to find the expansion of the $P$-dependent term in an arbitrary classical background.  

The terms $\delta_\phi T_{nn}$ and $\delta_{\kappa h} T_{nn}$ in \eqref{Hamexp} vanish in vacuum, but not with a nonzero background $\phi_0$.  They can be eliminated by passing to ``perturbation picture"\cite{SE2d} or absorbed in $T^Q$.  A matter background also leads to $h\phi$ and $hh$ terms in $T^Q$.  We will defer treatment of such a nontrivial background for future work and focus on the vacuum case.

In the vacuum case, working about a solution ${ \calc}_n= 0$, the preceding expansions then give
\beq
{\tilde \calc}_n = \sqrt q\left(- \frac{2}{\kappa} L^{ij} h_{ij}  +\frac{2}{\kappa} \calp^{ij}h_{ij} - \frac{2}{\kappa\sqrt q}  K_{ij} p^{ij} 
+T^Q_{nn} + \breve t_{nn}\right)
\eeq
where $K_{ij}$ is the extrinsic curvature of the slices in the background solution, related to the background $P_{ij}$ by \eqref{conjmom}, and 
\beq
 \calp^{ij} = \frac{\kappa^4}{2q}\left[P^{ik}P_k^j - \frac{PP^{ij}}{D-2} - \frac{q^{ij}}{2}\left(P^{kl}P_{kl} -\frac{P^2}{D-2}\right)\right]\ ;
 \eeq
 recall that the classical $P^{ij}$ is $\calo(\kappa^{-2})$, so $\calp^{ij}$ is $\calo(\kappa^0)$.

Expansion of the constraint \eqref{momcons} is handled similarly, giving
\beq
{\tilde \calc}_i = \calc_i + \Pi_0\partial_i\phi + \Pi\partial_i\phi_0  -2\kappa h_{ij} D_k P^{jk} -2q_{ij} (\delta_{\kappa h} D_k) P^{jk} -\frac{2}{\kappa} q_{ij} D_k p^{jk} +  \Pi \partial_i\phi +\sqrt q \breve t_{ni}\ 
\eeq
where again the quadratic and higher-order terms in $h,p$ have been collected in $\breve t_{ni}$.
Again restricting to the vacuum case, $\phi_0=\Pi_0=0$, and using the statement that the background solves the constraints, this becomes
\beq
{\tilde \calc}_i= -\frac{2}{\kappa} q_{ij} D_k p^{jk} - 2 q_{ij} (\delta_{\kappa h} D_k)P^{jk} -2 \kappa h_{ij} D_k P^{jk} + \sqrt q (T^Q_{ni} +  \breve t_{ni} )\ .
\eeq
We will collect the terms linear in $h$ by defining a linear differential operator $\cal Q$ by
\beq
{\cal Q}_i^{jk} h_{jk} = \kappa q_{ij} (\delta_{\kappa h} D_k)P^{jk} + \kappa^2 D_kP^{jk} h_{ij}\ .
\eeq

Working about a classical background with $\calc_n=\calc_i=0$, the $\kappa\rightarrow0$ limit of the constraints gives the linear homogeneous equations
\beq\label{hamzero}
L^{ij} h_{ij}  - \calp^{ij}h_{ij}+ \frac{K_{ij}}{\sqrt q} p^{ij} =0
\eeq
and
\beq\label{momzero}
D_j p_i^{j} + {\cal Q}_i^{jk} h_{jk}=0\ ,
\eeq
constraining linearized perturbations $(h_{ij}, p^{ij})$ about the solution, {\it i.e.} linearized gravitational waves.  These are evolved by a quadratic hamiltonian, which may be found from the rightmost expression in \eqref{hreln}, and which is expected to give evolution similar to that for other quantum fields, {\it e.g.} as treated in \cite{GiPe}, such as Hawking production in a black hole background, {\it etc.}  

At nonzero $\kappa$, the constraints become
\beq\label{hamone}
L^{ij} h_{ij}  - \calp^{ij}h_{ij}+ \frac{K_{ij}}{\sqrt q} p^{ij} =\frac{\kappa}{2}\left(T^Q_{nn} + \breve t_{nn}\right)
\eeq
and
\beq\label{momone}
D_j p_i^{j} + {\cal Q}_i^{jk} h_{jk}=\frac{\kappa}{2}\sqrt{ q}\left(T^Q_{ni} + \breve t_{ni}\right)\  .
\eeq
In the classical theory the corresponding equations determine the perturbative ``Coulomb fields" induced by matter, and at higher orders also incorporate the nonlinearities resulting from gravitational energy.  At leading order in $\kappa$, the solutions to \eqref{hamone}, \eqref{momone} are of course highly nonunique, since a solution of the homogeneous equations \eqref{hamzero}, \eqref{momzero} may be added to any given solution.

\section{Leading perturbative dressing}

In the quantum theory, finding gauge-invariant operators $O$ that commute with the constraints, \eqref{conscomm}, can be approached by perturbatively solving for operators that commute with \eqref{hamone}, \eqref{momone}.  Solutions can be found, beginning with an operator of the quantum field theory to which we couple gravity.  This is done by gravitationally dressing that operator, as has been described to leading order in perturbation theory about flat space in \cite{SGAlg}\cite{DoGi1}\cite{DoGi2,DoGi3}\cite{QGQF}\cite{DoGi4,SGsplit} and about anti de Sitter space in \cite{GiKi}.  Here we will extend those constructions to a more general background.

This gravitational dressing is most easily studied in the situation where the background satisfies $P^{ij}=0$, corresponding to vanishing extrinsic curvature of the time slices of the background metric.  This includes the case of flat and AdS backgrounds with standard time slicings.  However, we would also like to consider evolution that for example perturbs about black hole solutions, either with flat or AdS asymptotics.  For example in the case of the Schwarzschild solution, one may consider a general stationary slicing  that is spherically symmetric\cite{NVU}\cite{SEHS}\cite{GiPe}, 
\beq\label{sliceeq}
x^+=t+S(r)\ ,
\eeq
specified by a slicing function $S(r)$,
 where $x^+$ is the ingoing Eddington-Finkelstein coordinate.  The nontrivial components of the extrinsic curvature of the slices are then given by
\beq
D_rN_r = \partial_r N_r -\gamma^r_{rr} N_r\quad, \quad D_\theta N_\theta = -\gamma^r_{\theta\theta} N_r\ ,
\eeq
and so vanish if and only if $N_r=0$.  The expression\cite{NVU} $N_r=1-fS'$, with $-f$ the coefficient of $dx^{+2}$ (see below), then implies this is true only for $S'=1/f$, which is the case of Schwarzschild time slices.  These lead to a singular basis for perturbations at the horizon, and as explained in \cite{GiPe} this can be avoided with a more general choice of slices.  But, this therefore requires considering $P^{ij}\neq0$; similar statements hold for the case of black holes in AdS.

The construction of \cite{SGAlg}\cite{DoGi1}\cite{DoGi2,DoGi3}\cite{QGQF}\cite{DoGi4,SGsplit} writes the linear order dressing of an underlying QFT operator $O_0$ as
\beq\label{dressop}
O = e^{i\int d^{D-1}x \sqrt{q} V^\mu(x)(T_{n\mu} + \breve t_{n\mu})} O_0 e^{-i\int d^{D-1}x \sqrt{q} V^\mu(x)(T_{n\mu} + \breve t_{n\mu})}\ ;
\eeq
as long as we work to linear order in $\kappa$ the exponential is not strictly necessary, but is convenient and suggestive.  (To leading order about a vacuum solution $T^Q$ of \eqref{Hamexp} simplifies to $T$ of matter perturbations; we have also included the stress tensor $\breve t_{n\mu}$ for metric perturbations, in anticipation of the possibility that $O_0$ could also include such perturbations.)  Here the dressing functions $V^\mu(x)$ are functionals of the metric perturbation which are fixed by the condition that the dressed operator $O$ commute with the constraints.\footnote{Papers by Fr\" ob, Lima, and collaborators\cite{Frob:2017apy,Frob:2017lnt,Frob:2017gyj,Frob:2018tdd,Frob:2022ciq,Frob:2023awn,Frob:2023gng,Frob:2023vay} have also studied construction of leading-order  gravitationally-dressed observables.  The equivalence of their approach is seen, {\it e.g.} in the case of dressing of a scalar field, by noting that if our dressing $V^\mu$ defines a map $\chi$ through $\chi(y)= y + V(y)$ (Lorentz indices suppressed), then the map $X(x)$ given for example in (3) of \cite{Frob:2023gng} is $X(x) = \chi^{-1}(x)$.  Then the scalar version of (7) of that reference is the same form as $\phi(y+V(y))$ of ref.~\cite{DoGi1} eq.~(33), and the transformation properties under diffeomorphisms of $X(x)$ that they give follow from those of $V$ in \cite{DoGi1}.  This means that their dressing in eq. (8) corresponds to a special case of (39) of \cite{DoGi1}, up to a total derivative.  However, this difference is important, since their (8) does not transform correctly under harmonic diffeomorphisms.  The missing total derivative also appears to explain the claim of \cite{Frob:2022ciq,Frob:2023gng} that  they have observables commuting outside the light cone, in contradiction to the generic noncommutativity found in \cite{DoGi1} and to the dressing theorem of\cite{DoGi2}.}

\subsection{Vanishing background extrinsic curvature: $P^{ij}=0$}

We first consider this simplifying case. Specifically, generalizing the flat space construction\cite{DoGi1,SGsplit}, we anticipate that $V^n(x)$ takes the form
\beq
V^n(x) = -\frac{\kappa}{2} L^{-1}_{ij} p^{ij} = \frac{\kappa}{2} \int d^{D-1} x'  \check h_{ij}(x',x) p^{ij}(x')\ ,
\eeq
where the inverse $L^{-1}$ is given by a Green's function solution to
\beq
L_{x'}^{ij}  \check h_{ij}(x',x) = -\frac{\delta^{D-1}(x'-x)}{\sqrt q}\ .
\eeq
These Green's functions are highly non-unique, corresponding to the non-uniqueness of the perturbative classical solutions.  Explicit examples of this nonuniqueness in dressings have been described for perturbations about a flat background\cite{DoGi1}.  Examples there include explicit expressions describing either  line-like or Coulomb-like gravitational fields\cite{DoGi1}, and generalize to a broad class of instantaneous configurations of the field.  This nonuniqueness extends here to more general backgrounds, and again corresponds to differences by homogeneous solutions, corresponding to source-free gravitational waves.\footnote{This also means, as also described in \cite{DoGi3,DoGi4,SGsplit} that soft charges are largely decorrelated with the quantum state of matter in a region.  For the most part they depend on the arbitrary choice of gravitational dressing (which may be specified {\it e.g.} through imposition of boundary conditions), with the only necessary correlation through the total Poincar\'e charges of the matter state.}
With such a Green function, the commutator with the constraint $\tilde \calc_n$ is easily seen to give
\beq\label{Vncom}
[\tilde \calc_n(x),V^n(x')] = i \delta^{D-1}(x'-x) + \calo(\kappa)\ .
\eeq
As a consequence, using the expression \eqref{dressop} and assuming the $V_i$ term doesn't contribute (see below),
\beq 
[\tilde \calc_n(x),O] = 0+\calo(\kappa)\ ,
\eeq
with the commutator of the leading-order term from $(T+\breve t)$ in $\tilde \calc_n$ being cancelled by the term arising from \eqref{Vncom}.\footnote{Note that the leading order (in $\kappa$) transformation of the metric perturbations leads to terms that cancel the leading non-invariance of the operator $\calo_0$ in \eqref{dressop}.  Higher-order terms in the transformation of the metric perturbation then contribute to the $\calo(\kappa)$ terms here.}

To solve the momentum constraint \eqref{momone}, we consider dressing functions of the general form
\beq
V^i(x)=\kappa\int d^{D-1} x' G^{ijk}(x',x) h_{jk}(x')\ ,
\eeq
and seek a solution of the equation
\beq\label{Vicom}
[\tilde \calc_i(x), V^j(x')] = i \delta^j_i \delta^{D-1}(x'-x) +\calo(\kappa)\ .
\eeq
From the canonical commutators \eqref{pertcom}, we find this holds if
\beq
2 q_{ij} D_k G^{ljk}(x,x') = \delta _i^l\delta^{D-1}(x-x')\ .
\eeq
In a flat background, solutions are given by\cite{SGsplit}
\beq
V^i(x)=\int d^{D-1} x' \check h^{jk}(x',x) \gamma^i_{jk}\ .
\eeq
In more general backgrounds, the solutions are seen to correspond to Green functions for the equations for linearized metric perturbations, \eqref{hamone}, \eqref{momone}, and therefore should exist once boundary conditions are specified to fix a specific solution.  
One can also easily check that 
\beq\label{crosscom}
[\tilde \calc_n(x),V^i(x')]=\calo(\kappa)\quad ,\quad [ \tilde \calc_j(x), V^n(x')] =\calo(\kappa)\ .
\eeq
Then, given the commutators \eqref{Vncom}, \eqref{Vicom}, \eqref{crosscom}, we find that the constraints have been solved to leading nontrivial order in $\kappa$ by the dressed operators \eqref{dressop},
\beq
[\tilde \calc_\mu(x),O]= 0+\calo(\kappa)\ .
\eeq

\subsection{$P^{ij}\neq 0$}

Once one sees this structure, it is apparent how one can generalize to the case of background $P^{ij}\neq 0$.  We now define
\beq\label{genvn}
V^n(x)=\frac{\kappa}{2} \int d^{D-1}x' \left[ \check h_{ij}(x',x) p^{ij}(x') - \check p^{ij}(x',x) h_{ij}(x')\right]
\eeq
and
\beq\label{genvi}
V^i(x)= \kappa \int d^{D-1}x' \left[ G^{ijk}(x',x)h_{jk}(x') + H^i_{jk}(x',x) p^{jk}(x')\right]\ 
\eeq
where $\check h_{ij}$, $\check p^{ij}$, $G^{ijk}$, and $H^i_{jk}$ are c-number functions.
Then, the hamiltonian constraint gives
\beq
[\tilde \calc_n(x), V^n(x')] = -i\sqrt q \left(L^{ij}-\calp^{ij}\right)\check h_{ij}(x,x') - i K_{ij} \check p^{ij}(x,x') + \calo(\kappa)\ ;
\eeq
requiring this commutator to be of the form \eqref{Vncom} then gives
\beq
 \left(L^{ij}-\calp^{ij}\right)\check h_{ij}(x,x') +\frac{K_{ij}}{\sqrt q} \check p^{ij}(x,x') = -\frac{\delta^{D-1}(x-x')}{\sqrt q} \ .
 \eeq
 Leading order vanishing of the commutator of $V^n$ with the momentum constraint likewise gives
 \beq
 D_j\check p_i^j(x,x') + {\cal Q}_i^{jk} \check h_{jk}(x,x') =0  \ .
 \eeq
 This generalizes the above Green function problem, and so should again have solutions for $\check h_{ij}$ and $\check p^{ij}$ by the relation to the classical problem of finding linearized solutions on the background.
 
 Requiring the correct leading order commutators of the constraints with $V^i(x)$, \eqref{Vicom} and \eqref{crosscom}, likewise gives the equations
\beq
2q_{ik} D_l G^{jkl}(x,x') - 2{\cal Q}_i^{kl}H^j_{kl}(x,x') = \delta _i^j\delta^{D-1}(x-x')
\eeq
and
\beq
 \left(L^{jk}-\calp^{jk}\right)H^i_{jk}(x,x') -\frac{K_{jk}}{\sqrt q} G^{ijk}(x,x')=0\ ,
 \eeq
 which again corresponds to a Green function problem for linearized perturbations.  Once $\check h_{ij}$, $\check p^{ij}$, $G^{ijk}$, and $H^i_{jk}$ have been determined by solving these equations, together with specification of the homogenous part of the solution {\it e.g.} through boundary conditions, then 
\eqref{genvn} and \eqref{genvi}, together with \eqref{dressop}, give the dressed operator $O$ to leading nontrivial order in $\kappa$.

Significant features of the role of gauge invariance can be seen from the leading order in $\kappa$ construction of the dressed operators  \eqref{dressop} given here.  For example, the dressing modifies the commutators from those of the underlying LQFT operators $O_0$, such that operators associated with spacelike-separated regions generically no longer commute; examples can be given extending the discussion of \cite{DoGi1}.  Of course, to further understand the role and structure of the constraints and dressing, 
one would like to go beyond to higher orders in $\kappa$.  One does expect further difficulties here, in particular associated to infinities and the need to regulate operators.  We will leave further discussion of higher orders for future work, but will discuss some general features that already become apparent with these leading-order results.

\section{Description of evolution}

\subsection{General structure}

It is important to understand the general structure of evolution in quantum gravity, given the constraints of gauge invariance.  The leading-order construction of gauge invariant operators, and the more general structure of the hamiltonian and constraints, already appear to provide significant guidance to this structure.

In particular, we have given a leading-order construction of gauge invariant observables $O$, commuting with the constraints \eqref{conscomm}.  These then lead to states that evolve via the hamiltonian \eqref{hreln} of quantized general relativity, {\it e.g.} of the form
\beq\label{dressstate}
|\psi\rangle = O|0\rangle\ .  
\eeq
One can alternately construct dressed states directly from undressed states,
\beq
|\psi\rangle= e^{i\int d^{D-1}x \sqrt{q} V^\mu(x)(T_{n\mu} + \breve t_{n\mu})}|\psi_0\rangle\ .
\eeq
As a simple example, one could begin with the basic scalar field operator, $O_0=\phi(x)$, and then construct the corresponding dressed operator $O$, given to leading order by \eqref{dressop}, or equivalently\cite{DoGi1} by $\Phi(x)=\phi(x^\mu + V^\mu(x))$.  The resulting operator can be thought of as creating from the vacuum a quantum of the field $\phi$, together with its corresponding gravitational field.  As we have emphasized, the gravitational part of the operator is non-unique, corresponding to the fact that there are different possible gravitational field configurations dressing the particle, differing at leading order by free gravitational excitations.

Evolution can be thought of in two ways, related through the two expressions for the hamiltonian \eqref{hreln}.  Since the rightmost expression there is of the standard form of a LQFT hamiltonian, it determines evolution of the state by telling us how the $\phi$ and gravitational parts of the state evolve like standard quantum fields.  This evolution is of course gauge dependent, through its dependence on $N$ and $N^i$.  We expect it to correspond to the quantum evolution of the matter state created by $O$, together with the quantum gravitational field created by the gravitational piece of the operator.

Or, one can describe evolution in terms of the middle expression in \eqref{hreln} which is written in terms of the constraints.  The boundary hamiltonian $H_\partial$ contributes to the time dependence of the state, because the gravitational dressing generically extends into the asymptotic region\cite{DoGi2,GiKi}.  One might have anticipated that the constraints $\calc_\mu(x)$ annihilate the state, so that this is the only time dependence.  This would incorporate the statement that the state ``satisfies the Wheeler-DeWitt equation," since the constraint $\calc_n(x)$ corresponds to the Wheeler-DeWitt operator.  However, we have found that this would be inconsistent with the basic commutators and in particular with evolution such as described by the Heisenberg equations \eqref{qeqn}, \eqref{Peqn}.  Instead the state is annihilated by ``half" of the constraints  (and of the Wheeler-DeWitt operator), \eqref{physstate}.  This implies that the constraint terms in \eqref{hreln} also contribute to gauge-dependent evolution of the state, though as we have argued above not to evolution of matrix elements of gauge-invariant operators.  Alternately, transition amplitudes of the form 
\beq
\langle\psi'| e^{-iHt}|\psi\rangle 
\eeq
will also exhibit time dependence.  

\subsection{Bubble evolution, cosmology, and field-relational observables}

This raises the question of the description of evolution on slices that coincide at infinity, but not in a region in the interior of spacetime, so that the asymptotic lapse and shift vanish, implying $H_\partial=0$.  Then, the full hamiltonian commutes with the gauge-invariant operators $O$.  This suggests that their evolution is trivial in such ``bubble" evolution\cite{Tomo1946,Kuch-bubble}.  This is also the case for closed cosmologies, with no boundary term.  In both of these cases the hamiltonian is typically explicitly time-dependent, with additional subtleties\cite{ToVa,CMOV,CFMM,AgAs,MuOe}.  Once again, we can anticipate that the physical states are annihilated by ``half" of the constraints.  An alternate way to then describe evolution is in terms of a different kind of relational observable that is not gravitationally dressed; an example of such a field-relational observable is
\beq
\int d^Dx \sqrt{|g|} O_0(x) f(Z^I(x))
\eeq
where $Z^I$ are $D$ dynamical ``locator" fields and $f(Z^I)$ is chosen so that in a particular state for these fields its support is localized near a particular point.  An example is using the value of the inflaton field in inflation to localize in time to the reheating time; 
for further discussion (including of limitations to localization) see  \cite{GMH}.  We leave further exploration of such evolution for future work.  

\subsection{Other specific examples}
\label{exevol}

Beyond a flat space background, it is of interest to better understand gravitational evolution in other backgrounds such as those of black holes, AdS, or black holes in AdS.  The static cases can be subsumed in the line element
\beq
ds^2=-f(r) dx^{+2} + 2 dx^+ dr + r^2 d\Omega^2
\eeq
where
\beq
f(r)=1+\frac{r^2}{R_\Lambda^2} - \left(\frac{R_0}{r}\right)^{D-3}\ ,
\eeq
$R_\Lambda$ is the AdS radius, and 
\beq
R_0^{D-3} = \frac{16\pi G_D M}{(D-2) A_{D-2}}
\eeq
with $A_{D-2}$ the area of the unit sphere.  Then, introducing a stationary slicing \eqref{sliceeq} given by a slice function $S(r)$ yields the ADM background solution\cite{NVU}
\beq
N^2=\frac{1}{S'(2-fS')}\quad,\quad N_r= 1-fS' \quad,\quad q_{rr}= S'(2-fS')\quad,\quad
\eeq 
and with angular components the standard round metric of radius $r$.  

Evolution may then be described perturbatively about this solution, using the preceding general construction.  Specifically, we may consider a dressed state \eqref{dressstate} on a slice taken to be an initial slice.  The evolution of this state can be described via either of the forms of the hamiltonian \eqref{hreln}.  The latter form in particular gives a standard description of field evolution, and so evolves the matter perturbation together with the perturbative gravitational field corresponding to its dressing in standard field theory fashion.\footnote{While Lorentzian evolution for black holes has been considered previously, see {\it e.g.} \cite{CaTe}, the treatment outlined here extends to more general slicings than Schwarzschild, and including the black hole interior.}

In this way, one for example finds a perturbative expression for the bulk hamiltonian for an AdS black hole to leading order in $\kappa$, or in the language of the AdS/CFT correspondence, in $1/N$, also making connection with the discussion in \cite{GiPe} of this approach to defining such a hamiltonian.  
Alternately, by virtue of the constraints, the hamiltonian is related to a boundary hamiltonian as in \eqref{hreln}.
We  defer more detailed investigation of this evolution to future work.  

\section{Conclusion and directions}

In conclusion, we have shown that, starting from an ADM parameterization of the geometry and the corresponding construction of the hamiltonian, leading order perturbative gravitational states may be constructed and their evolution described.  The states and evolution have a gauge symmetry, generated by the constraints, and perturbative solution of the constraints to construct gauge-invariant operators and states can be accomplished by gravitational dressing operators of an underlying field theory.  Such a construction has been found to leading perturbative order about a general background, in terms of certain generalized Green functions of the given background, generalizing earlier constructions in a flat background\cite{SGAlg}\cite{DoGi1}\cite{DoGi2,DoGi3}\cite{QGQF}\cite{DoGi4,SGsplit}.  The resulting gravitational part of the state is not uniquely determined, since it can be changed by addition of a piece corresponding to an arbitrary source-free propagating gravitational wave. The state gotten by acting with such a dressed operator on a vacuum state is then evolved by the hamiltonian, which may be described as a standard local QFT hamiltonian including the spin-two perturbative gravitational field, or alternately may be written in terms of a boundary hamiltonian, up to terms proportional to the constraints.

There are multiple directions for further work.  Within the framework of local QFT, as noted above, we would like to better understand bubble evolution, on slices that match at infinity, and the related description of cosmological evolution, and the connection to other field-relational observables that are more useful in that context. There are also related issues that occur when the background slicing has an explicit time dependence\cite{ToVa,CMOV,CFMM,AgAs,MuOe}, which deserve to be more closely investigated.   It also seems useful to have a more complete description of the evolving perturbative state of black holes, whether or not in AdS, {\it e.g.} generalizing \cite{GiPe} to include gravitational dynamics.  And, the problem of solving the constraints connects directly to a leading argument for the origin of holography \cite{Maroholo,Maroholonost,Jacoholo,SGholo}, which is important to better understand.  This also connects to the question of in what sense information can be localized in a gravitational theory, either because of the argued existence\cite{Maroholo,Maroholonost} of a holographic map, or a similar argument\cite{DoGi3} that states internal to AdS may be observed by boundary observables, {\it if} the constraints are solved.\footnote{Such information has even been argued to be accessible perturbatively\cite{Rajuetal}, although only if one can measure exponentially small quantities\cite{SGsub}.} This connects directly to the question of the extent to which subsystems may be defined\cite{QFG,QGQF,SGsub,SGsnow} in gravitational theories, whether exactly or approximately.  We note that a perturbative description of the evolution like we have outlined is consistent with the perturbative solution of the constraints, and appears to describe a black hole with a growing number of  internal states entangled with the exterior, and corresponding missing information if the black hole disappears at the end of evolution.  
Thus, while the perturbative solution of the constraints via the dressing does appear to provide some additional sensitivity to the black hole state, it does not appear to resolve the unitarity problem, in contrast to recent claims\cite{Rajuetal}, and in particular does not obviously provide a mechanism for transfer of information from the black hole.

In clarifying these issues, understanding better the structure of higher-order solution to the constraints seems important (and it seems important to clarify  the challenges to finding higher-order solutions).  We would also like to better understand the structure of the algebras associated to dressed observables.  It can be observed that the leading-order perturbative dressed observables, \eqref{dressop} and related expressions in \cite{DoGi4}, have similar structure to the observables in the crossed-product construction, argued\cite{Wittcross}\cite{CLPW} to convert type III von Neuman algebras into type II.  Of course, the noncommutativity of perturbative observables associated with different regions\cite{SGAlg,DoGi1}, due to the dressing, appears to be a likewise important modification of the underlying field theory structure; further investigation of these questions is in progress.

The modification of local algebras in gravity illustrate the general statement that locality is remarkably subtle in theories with structures like gauge symmetries.  In QED or standard nonabelian gauge theories, observables found by dressing underlying matter operators are generically nonlocal, as with the gravitationally dressed operators discussed above.  Put differently, the problem of solving the constraints is generically a nonlocal one, with a nonlocal solution.  However, in gauge theories based on an internal group, locality is still realized since there also exist gauge-invariant operators that are local, such as Wilson loops or other neutral observables confined to a neighborhood.  In gravity, the gauge symmetry is that of transformations including Poincar\'e symmetries; any local observable thus carries nontrivial charge, and so is nonlocal when its gravitational dressing is included\cite{DoGi2}.  
In short, gauge theories with an internal symmetry appear to be barely local, but it is less clear what locality properties quantum gravity has.  

There are also important directions that appear to go beyond the framework of local QFT.  For one thing, if we consider evolution of a black hole, like that described in \ref{exevol}, that appears to lead to the breakdown of unitarity, also noted above, associated with the ``black hole information paradox" or ``unitarity crisis."  This can be encapsulated in a ``black hole theorem\cite{SGthm}": unitary evolution, and the statements that black holes behave like subsystems, that field configurations outside them evolve independently of the black hole internal state, and that they disappear at the end of their evolution, are inconsistent.  
We expect modification of both the structure of the Hilbert space and the hamiltonian as compared to those given by quantization of GR like that we have described.  In a more complete description,  we expect that the fundamental quantum variables are likely {\it not} those of fields moving on a background metric, with states labelled such as  $|q_{ij}(\cdot),\phi(\cdot)\rangle$, but that these variables only give an approximate description of the states.  An important question is what is the more accurate and complete description of the variables parameterizing the wavefunction.  This, then, closely relates to the question of what are the fundamental observables, and the ultimate form of the hamiltonian and its interactions, as well as the symmetries of the theory. 

If in a more complete description of the Hilbert space black holes still effectively behave as subsystems, the ``black hole theorem" tells us that unitarity apparently requires interactions that go beyond the quantized GR/local QFT description, and specifically such that the evolution of the black hole exterior depends on the black hole internal state.  An approach to parameterizing such interactions has been developed in \cite{SGmodels,BHQIUE,NVNL}\cite{NVU,BHQU}.  The present work serves as an even firmer foundation on which to describe their parameterization, if they can be regarded as   corrections to the evolution governed by local QFT plus quantized GR.  
In short, one can describe corrections $\Delta H$ to the hamiltonian of \eqref{hreln}, constrain their properties, and investigate their possible observational effects for example in electromagnetic or gravitational wave observations of the near-horizon regime\cite{BHQU}.  The contributions to $\Delta H$ are plausibly small both far from the black hole and even in the near horizon region, but of course are expected not to be small in the deep black hole interior.  
More systematic analysis is planned for future work.  

 It is believed in a large segment of the quantum gravity community that a fundamental description like this may arise from a dual large-$N$ gauge theory; if this is true, it is important to understand and characterize the departures from the bulk LQFT description that this implies, for example as corrections to the hamiltonian \eqref{hreln}.  But, it seems quite likely that the more fundamental description arises in connection with some other mathematical structure on Hilbert space\cite{QFG,QGQF}, which it is our goal to infer and further describe.

\vskip.3in
\noindent{\bf Acknowledgements} 

This material is based upon work supported in part by the U.S. Department of Energy, Office of Science, under Award Number {DE-SC}0011702, and by Heising-Simons Foundation grant \#2021-2819.   We thank Z. Wang for useful comments.  Work to complete this paper was carried out at the Aspen Center for Physics, which is supported by National Science Foundation grant PHY-1607611.

We dedicate this paper to the memory of Jim Hartle, an esteemed colleague, valued mentor, good friend, and general  inspiration.

\appendix
\section{Gauge-invariant canonical quantization of electromagnetism}
\label{appa}

This appendix will illustrate aspects of the quantization of a gauge-invariant theory in the simpler context of QED, with particular focus on the ``gauge-invariant canonical quantization" used for gravity in the main text.  

The starting point is the gauge-invariant lagrangian
\beq\label{QEDact}
\call = -\frac{1}{4} F_{\mu\nu} F^{\mu\nu} + \call_m\ ,
\eeq
with $F_{\mu\nu} = \partial_\mu A_\nu - \partial_\nu A_\mu$ and $\call_m$ a matter lagrangian, for example that coupling to a fermion field,
\beq
\call_m = i\bar \psi (\partial_\mu + i e A_\mu)\gamma^\mu\psi - m \bar \psi\psi\ .
\eeq
The momentum conjugate to $A$ is
\beq
\pi^\mu=\frac{\partial \call}{\partial \dot A_\mu} =-F^{0\mu}\ .
\eeq
As a result $\pi^0=0$.  This can be implemented working with a reduced phase space, where $A_0$ is no longer treated as a canonical variable.\footnote{For certain gauges, additional care is needed here; an example is axial gauge, $A_z=0$, as is further explored in {\it e.g.} \cite{DoGi1}.}
Spatial components of the momenta are the electric field,
\beq
\pi^i= \partial_0 A^i + \partial^i A^0 = -E^i\ .
\eeq
Then the hamiltonian form of the Maxwell part of the action \eqref{QEDact} becomes
\beq
S=\int d^4x\left(\pi^i\dot A_i -\calh\right)\ ,
\eeq
with Maxwell hamiltonian
\beq\label{Maxham}
H=\int  d^3x \calh=   \int d^3x\left( \frac{E^{i2}}{2}+ \frac{B^{i2}}{2} +E^i\partial^i A^0\right)\ 
\eeq
and $B_i=\epsilon_{ijk}F^{jk}/2$.  A constraint arises from varying $A^0$, which behaves like a Lagrange multiplier, giving
\beq\label{Econst}
\partial_i E^i = j^0
\eeq
where we have included the contribution from the matter current.  $A^0$ remains unfixed, and is taken to be arbitrary.

As described in the main text, there are different options for how to treat quantization: Dirac, covariant gauge ``fixing" (breaking), BRST, refined algebraic, and what we will call gauge invariant canonical quantization, and examine here.
The starting point for this is the canonical commutators,
\beq\label{CCRs}
[\pi^i(x,t),A_j(x',t)]= -[E^i(x,t),A_j(x',t)]= -i\delta^i_j \delta^3(x-x')\ ,
\eeq
Then, the constraint \eqref{Econst} generates the gauge transformations,
\beq\label{EMgauge}
[\partial_iE^i(x), A_j(x')]=i\partial_j \delta^3(x-x')\ .
\eeq
$A_0$ remains an arbitrary c-number function, which also behaves like a gauge parameter; the condition $\pi^0=0$ is implemented through 
independence of physical quantities on $A_0$. 
Eq.~\eqref{EMgauge} also shows that vanishing of the constraint ({\it e.g.} consider $j^0=0$) on the Hilbert space would imply $\langle0|[\partial_iE^i,A_j]|0\rangle=0$ and
be incompatible with the basic commutators, 
unless commutators are modified as in the Dirac approach.

The canonical commutation relations \eqref{CCRs} can be represented in the usual fashion in terms of orthonormal polarization vectors $\epsilon_{i\lambda}(k)$ and annihilation/creation operators $a_{k\lambda}/a^\dagger_{k\lambda}$ as
\bea
A_i(x,0)&=&\sum_\lambda \int \widetilde{dk}\left[\epsilon_{i\lambda}(k) a_{k\lambda} e^{ikx} + h.c. \right] + a_i(x)\cr
E_i(x,0)&=&\sum_\lambda \int \widetilde{dk}\left[ ik \epsilon_{i\lambda}(k) a_{k\lambda} e^{ikx} + h.c. \right] \ ,
\eea
with 
\beq
[a_{k\lambda},a^\dagger_{k'\lambda'} ]= (2\pi)^3 2|k| \delta_{\lambda\lambda'}\delta^3(k-k')\ ,
\eeq
$\widetilde {dk}= d^3k/(2\pi)^3 2|k|$, and $a_i(x)$ a c-number function arising from gauge invariance.  General states can be constructed in the form $\prod(a^\dagger_{\kappa\lambda})|0\rangle$.  However, the constraints (focussing on the free theory) are implemented as a physical state condition
\beq\label{statecond}
\partial_i E^{i+}|\psi\rangle=0\ 
\eeq
in terms of the annihilation piece of $\partial_i E^i$, corresponding to $a_{3k}|\psi\rangle=0$ in a standard choice of basis.

Evolution can be studied in Schr\"odinger or Heisenberg pictures.  In the former, the evolution with $H$,
\beq
i\partial_t|\psi\rangle = H|\psi\rangle\ ,
\eeq
depends on the arbitrary $A_0$; for $A_0$ of compact support this results in a gauge-dependent piece of the evolution of the state,
\beq
\delta_{A_0} |\psi\rangle=i\delta t \int d^3x \partial_i E^{i-}(x) A_0(x)\,  |\psi\rangle\ .
\eeq
However, this piece is orthogonal to another physical state $|\psi'\rangle$, by \eqref{statecond}.  Moreover, consider evolution of the matrix element of an operator depending on the canonical variables $E_i$, $A_i$, but not on $A_0$,
\beq\label{exev}
\partial_t\langle\psi'|O|\psi\rangle = i\langle \psi'| [H,O]|\psi\rangle\ .
\eeq
Gauge invariance of $O$ also requires $[\partial_iE^i,O]=0$, and in that case only the gauge-invariant ($A^0$ independent) part of $H$ contributes to the evolution  \eqref{exev}: evolution of matrix elements of such gauge-invariant operators is gauge invariant.  In contrast, $\partial_t \langle\psi'|A_i|\psi\rangle$ is not gauge invariant.

Equivalently, evolution can likewise be described by converting to Heisenberg picture.  The Heisenberg equations are
\beq
\partial_tE_i = i[H,E_i] = \nabla\times B_i - j_i\ ,
\eeq
and
\beq
\partial_t A_i= i[H,A_i]=-E_i +\partial_i A_0\ 
\eeq
These are supplemented by the constraint \eqref{Econst}, which as we have seen is {\it not} treated as an operator equation on physical states.  From these equations we find that the evolution $\partial_t E_i$ and also $\partial_t B_i$ are independent of the arbitrary gauge parameter $A_0$.  We can likewise consider gauge invariant operators built by dressing matter operators, and their evolution is also gauge-independent.  

\section{Gauge Transformations and diffeomorphisms}
\label{appb}

This appendix will discuss the role of the constraints in generating gauge transformations, in the canonical formalism used in the main text.  For simplicity we will consider gravity coupled to the scalar field with lagrangian \eqref{scalarlagrange}, which is also treated classically in  \cite{KuchII}.  Canonical data for the scalar field is the field $\phi$ and its canonical conjugate momentum $\Pi$. For the geometry, the phase space variables are  $D-1$-dimensional spatial metric $q_{ij}$ and the conjugate momentum $P^{ij}$, as well as the lapse and shift $N$, $N^i$, and their conjugate momenta.  However, the latter momenta vanish for the Einstein action, analogously to the vanishing of $\pi^0$ in QED (see preceding appendix).  As a result, one can commonly work on a reduced phase space where they are set to zero and where $N$ and $N^i$ are no longer treated as canonical variables, like with the electromagnetic case.\footnote{For further discussion of this see for example \cite{RTH} or \cite{PSS}.  Note that as with QED, additional care is needed when imposing certain gauges, such as ``axial" or Fefferman-Graham gauges, $h_{z\mu}=0$.}  We will then study the transformation generated by the general superposition of the constraints
\beq
C[\xi, \xi^i] = \int d^{D-1} x  \left(\xi \calc_n + \xi^i \calc_i\right)\ ,
\eeq
acting on the reduced phase space.
The explicit form of the constraints $\calc_n$, $\calc_i$ was given in eqs.~\eqref{hamcons}, \eqref{momcons}.

We begin by considering the action on matter.\footnote{For recent treatment of canonical quantization of matter on a general background, see \cite{GiPe}.}  
The canonical commutators are
\beq\label{scalarCanCom}
[\phi(x), \Pi(x')] = i \delta^{D-1}(x-x')\ ,
\eeq
where $\Pi$ is the densitized canonical momentum, satisfying
\beq
\label{SOM}
\Pi = \sqrt q \partial_n\phi\ .
\eeq
These commutators and the explicit form \eqref{hamcons}, \eqref{momcons} of the constraints give the commutator
\beq\label{scalarcom}
i[C[\xi, \xi^i],\phi] = \xi \frac{\Pi}{\sqrt q} + \xi^i\partial_i\phi \ .
\eeq
The second term is the action of a spatial diffeomorphism on $\phi$.  If we also use the 
 Heisenberg equation of motion \eqref{SOM}, we find that the full commutator becomes the  Lie derivative with respect to the vector $\xi^\mu$ with components
\beq
\xi^{\mu} = (\xi, \xi^i)\ ;
\eeq
explicitly
\beq\label{Cphicom}
i[C[\xi, \xi^i],\phi] = \pounds_{\xi^\mu} \phi\ + \xi\left(\frac{\Pi}{\sqrt q} -\partial_n\phi\right)\ ,
\eeq
and so the transformation generated by $C[\xi, \xi^i]$ can be identified as a general diffeomorphism for configurations satisfying the equations of motion.

One can likewise compute the commutator of the constraints with $\Pi$, which gives
\bea\label{Cpicom}
i [C[\xi, \xi^i],\Pi] &=& \partial_{i}(\xi q^{ij}\sqrt{q}\partial_j \phi) + \partial_{i}(\xi^{i}  \Pi)\\
&=& \partial_{\mu}(\xi n^{\mu}\Pi) +  \partial_{i}(\xi^{i}  \Pi) + \partial_{\mu}\left(\xi \sqrt{q} g^{\mu \nu}\partial_{\nu} \phi  \right)+ \partial_{\mu}[\xi n^{\mu}(\sqrt q\partial_n\phi-\Pi)]
\eea
Once again the second term gives a spatial diffeomorphism.  If in addition $\xi$ is identified with the lapse $N$, and the scalar field equations are satisfied,
$C[\xi, \xi^i]$ also generates the action of a diffeomorphism, $\pounds_{\xi^\mu} \Pi$.  In the case where $\xi^i$ is also taken to be the shift, eqs.~\eqref{Cphicom}, \eqref{Cpicom} also give the time derivative defined via  \eqref{xiton}.

The transformations of the spatial metric $q_{ij}$ and the conjugate momentum $P^{ij}$ are similar in structure to those of the matter fields: using the canonical commutation relation \eqref{cancom} and the equations of motion, $C[\xi,\xi^i]$ generates diffeomorphisms.
Beginning with the commutator of the metric, this results in an expression analogous to \eqref{scalarcom},
\bea\label{Cqcom}
i [C[\xi, \xi^i],q_{kl}] &=& \frac{\kappa^2}{\sqrt{q}}  \xi \left(P_{kl}-\frac{P q_{kl}}{D-2}\right)+D_k \xi_l+D_l \xi_k\\
&=& -2\xi K_{kl} +D_k \xi_l+D_l \xi_k + \xi \left[2 K_{kl} + \frac{\kappa^2}{\sqrt{q}}\left(P_{kl}-\frac{P q_{kl}}{D-2}\right)\right]\ .\label{CqcomLie}
\eea
The terms involving $\xi_i$ once again correspond to a spatial diffeomorphism.  
The last term, when set to zero, is the trace reverse of the relation of the conjugate momentum \eqref{conjmom} to the extrinsic curvature, which is a Heisenberg equation of motion in the canonical description. If this equation is satisfied, the RHS of \eqref{CqcomLie} is equal to
\beq\label{IndMetLie}
\pounds_{\xi^{\mu}} q_{kl} = \nabla_k \xi_l +\nabla_l \xi_k\ ,
\eeq
with the Lie derivative defined by using the $D$-dimensional expression $q_{\mu\nu}= g_{\mu\nu} + n_\mu n_\nu$ for the spatial metric.
This is the expected gauge transformation. If the vector $\xi^{\mu}$ is taken to be the time evolution vector \eqref{xiton}, the equation \eqref{Cqcom} gives the equation of motion for $q_{ij}$, and can be solved for the expression for the extrinsic curvature \eqref{excurv} given in the main text. 

Finally, for the ADM conjugate momentum, the commutator with the constraints gives
\bea\label{CPcom}
i [C[\xi, \xi^i],P^{kl}] &=& \frac{2\sqrt{q}}{\kappa^2}( D^{k} D^{l} \xi - q^{kl} D^2 \xi)- \frac{2\sqrt{q}}{\kappa^2} \xi \Big(R_q^{kl} - q^{kl} \frac{R_q}{2}\Big)\nonumber \\
&-& \frac{\kappa^2}{2\sqrt{q}} \xi \left[2 P^{ki} P_{i}^{l} -2 \frac{P^{kl} P}{D-2} - \frac{1}{2}q^{kl}\left(P^{ij} P_{ij}-\frac{P^2}{D-2}\right)\right]+ \frac{\sqrt{q}}{2} \xi S^{kl} \nonumber \\
&+& \partial_{i}(\xi^{i} P^{kl})- P^{ik}\partial_{i} \xi^{l}- P^{il}\partial_{i} \xi^{k} \ ,\label{momentumvar}
\eea
where the tensor $S_{\mu\nu} = q^{\lambda}_{\mu}q^{\sigma}_{\nu}T_{\lambda\sigma}[\Pi, \phi]$ is the projection of the scalar stress energy tensor \eqref{phistress}, written in terms of the canonical variables $\Pi$ and $\phi$, and has indices raised with the induced metric. Recalling that $P^{kl}$ is a tensor density, the final line of \eqref{momentumvar} is once again the action of a spatial diffeomorphism, $\pounds_{\xi^i}P^{kl}$. The relationship between the remaining terms proportional to $\xi$ and the normal component of the Lie derivative takes more work to illustrate. 
We begin by rewriting 
\bea\label{CPre}
i [C[\xi, \xi^i],P^{kl}] &=& \pounds_{\xi^{i}} P^{kl} + \Biggl\{\frac{2\sqrt{q}}{\kappa^2}( D^k D^l \xi - q^{kl} D^2 \xi)-\frac{2\sqrt{q}}{\kappa^2} \xi \Big(R_q^{kl} - q^{kl} \frac{R_q}{2}\Big)\nonumber \\
&-& \frac{\kappa^2}{2\sqrt{q}} \xi \left[2 P^{ki} P_i^l -2 \frac{P^{kl} P}{D-2} - \frac{1}{2}q^{kl}\left(P^{ij} P_{ij}-\frac{P^2}{D-2}\right)\right]+ \frac{\sqrt{q}}{2} \xi S^{kl}\Biggr\}\ .\label{commADMmom}
\eea
The Lie derivative in the normal direction can be defined by extending to tensors $P^{\mu\nu}$ and $K^{\mu\nu}$ on the full spacetime, with $K_{\mu\nu}=-q_\mu^\lambda q_\nu^\sigma \nabla _\lambda n_\sigma$.  
Then, when the Heisenberg equation \eqref{conjmom} (replacing the Latin with Greek indices) holds, one can show
\bea\label{LieCanMom}
 \pounds_{\xi n^{\lambda}} P^{\mu\nu} &=& \frac{2\sqrt{q}}{\kappa^2}\frac{\xi}{N}( D^{\mu} D^{\nu} N- q^{\mu\nu} D^2 N)-\frac{2\sqrt{q}}{\kappa^2} \xi \Big(R_q^{\mu\nu} - q^{\mu\nu} \frac{R_q}{2}\Big)\nonumber \\
&-& \frac{\kappa^2}{2\sqrt{q}} \xi \left[2 P^{\mu \lambda} P_{\lambda}^{\nu} -2 \frac{P^{\mu\nu} P}{D-2} - \frac{1}{2}q^{\mu\nu}\left(P^{\lambda\rho} P_{\lambda \rho}-\frac{P^2}{D-2}\right)\right]\nonumber \\
&+&\frac{2\sqrt{q}}{\kappa^2} \xi q^{\mu\lambda}q^{\nu\rho} \left(R_{\lambda\rho}-g_{\lambda\rho}\frac{R}{2} \right)\nonumber\\
&-& Nn^{\mu}P^{\nu\lambda}D_{\lambda}\left(\frac{\xi}{N}\right)-Nn^{\nu}P^{\mu\lambda}D_{\lambda}\left(\frac{\xi}{N}\right)\ , 
\eea
where we have also used the Gauss relation to simplify. Note that the first term is related to the acceleration, $a^{\mu} = n^{\nu}\nabla_{\nu} n^{\mu} = D^{\mu} \ln{N}$, and could be set to zero with the  choice of Gaussian normal coordinates. The second to last line of \eqref{LieCanMom} is proportional to the projected components of the $D$-dimensional Einstein tensor, and when the projected components of the Einstein equation hold can be replaced by $S^{\mu\nu}$. The last two terms have normal components to the surface.  When
 the equations of motion hold, and if we again take $\xi=N$, 
the term of \eqref{CPre} in braces matches the Lie derivative with respect to $Nn^\mu$, and so
the  RHS of \eqref{CPre} reduces to the Lie derivative with respect to $\xi^\mu$.
Additionally, the time evolution \eqref{Peqn} for the ADM conjugate momentum may be found from \eqref{commADMmom} if $\xi^{\mu}$ is taken to be given in terms of  the lapse and shift by \eqref{xiton}.

In conclusion the gauge transformations generated by the constraints acting on the reduced phase space variables $q_{ij}$, $P^{ij}$ correspond to the diffeomorphisms if the equations of motion hold, with $\xi$ identified as the lapse.

\mciteSetMidEndSepPunct{}{\ifmciteBstWouldAddEndPunct.\else\fi}{\relax}
\bibliographystyle{utphys}
\bibliography{gravevol}{}

\end{document}